\documentclass[12pt]{article}
\usepackage[a4paper,left=2cm,right=2cm,top=2cm,bottom=2cm]{geometry}       
\usepackage{cite}
\pdfoutput=1
\usepackage{graphicx}
\usepackage{bm}
\usepackage{amsmath,amssymb}
\usepackage{mathrsfs}
\usepackage{epstopdf}
\usepackage{authblk} 
\usepackage{IEEEtrantools}
\usepackage{hyperref}
\hypersetup{
	colorlinks=true,
	linkcolor=blue,
	filecolor=magenta,      
	urlcolor= cyan,
	citecolor= blue
}

\newcommand{\tf}{\tilde{f}}

\newcommand{\vp}{v_{+}}

\newcommand{\bw}{\bar{w}}

\newcommand{\ba}{\bar{\alpha}}
\newcommand{\bb}{\bar{\beta}}

\newcommand{\bd}{\bar{d}}

\newcommand{\tw}{\tilde{w}}

    \makeatletter
\def\@fnsymbol#1{\ensuremath{\ifcase#1\or \dagger\or \ddagger\or
		\mathsection\or \mathparagraph\or \|\or **\or \dagger\dagger
		\or \ddagger\ddagger \else\@ctrerr\fi}}
\makeatother

\title{Anti-Self-Dual Spacetimes, Gravitational Instantons and Knotted Zeros of the Weyl Tensor}
\author[1]{Snigdh Sabharwal\thanks{s.sabharwal@umail.leidenuniv.nl}}
\author[1]{Jan Willem Dalhuisen\thanks{Dalhuisen@physics.leidenuniv.nl}}
\affil[1]{Huygens-Kamerlingh Onnes Laboratory, Leiden University, P.O. Box 9504, 2300 RA Leiden, The Netherlands}
\date{}                                           

\begin{document}
\maketitle
\section{Abstract}
We derive a superpotential for null electromagnetic fields in which the field line structure is in the form of an arbitrary torus knot. These fields are shown to correspond to single copies of a class of anti-self-dual Kerr-Schild spacetimes containing the Sparling-Tod metric. This metric is the pure Weyl double copy of the electromagnetic Hopfion, and we show that the Eguchi-Hanson metric is a mixed Weyl double copy of this Hopfion and its conformally inverted state. We formulate two conditions for electromagnetic fields, generalizing torus knotted fields and linked optical vortices, that, via the zero rest mass equation for spin 1 and spin 2, defines solutions of linearized Einstein's equation possessing a Hopf fibration as the curves along which no stretching, compression or precession will occur. We report on numerical findings relating the stability of the linked and knotted zeros of the Weyl tensor and their relation to linked optical vortices. 

\section{Introduction}
Complex numbers enter general relativity in different ways. Be it in the form of complex transformations, for example the Newman-Janis method \cite{Adamo2014, Erbin2015}, to generate axisymmetric from spherically symmetric solutions of Einstein's equation, most notably the ones belonging to the Kerr-Schild (KS) spacetime \cite{Adamo2014, flaherty1976hermitian}, or in the study of complex spacetimes, where they find applications to various areas \cite{Held}. One of the main prospects of complex relativity is to be able to construct real exact solutions of Einstein's equation and while this seems to be a challenging problem, there have been some isolated results in \cite{Robinson1987, Robinson2002, Chudecki2018, Chudecki2018a, Plebanski1995a, Plebanski1998, Rozga1977}. A nice overview of complex relativity and it's different aims can be found in 
\cite{Mcintosh1985, Hall1985, Hickman1986a, Hickman1986, McIntosh1988}. Although the use of complex transformations and complex spacetimes for the purpose of generating new exact solutions of Einstein's equation is yet to find it's full potential, they have however been very successful in finding new solutions in flat spacetime theories. The generation of twist from twistfree Maxwell fields in flat space using a complex transformation can be traced to the works of A. Trautman \cite{Trautman1962} and Synge \cite{Synge}. Therein one finds the idea of the Robinson congruence, which was one of the motivations for twistor theory. Trautman's work on finding solutions to Maxwell's and Yang-Mills equations using Hopf fibrations \cite{Trautman1977} was also the antecedent to the theory of topological electromagnetism put forth by A. Ra\~{n}ada \cite{Ranada1989, Ranada1990}, where a topological non-trivial solution of source-free Maxwell's equations, an electromagnetic Hopfion, was found by means of the Hopf map. Presently, there have been many other approaches, better equipped with the construction of topological non-trivial Maxwell fields. An overview of the available methods can be found in \cite{Arrayas2017}.\\
\noindent In this article we describe a correspondence between a class of anti-self-dual (asd) solutions to Einstein's equation containing the Sparling-Tod metric \cite{Sparling1981, Tod1981} and a class of topological non-trivial exact solutions of free space Maxwell's equations by making use of the formalism developed by Pleba\'{n}ski \cite{Plebanski1975} and Tod \cite{Tod1982}. The field lines of the electromagnetic solutions are organized on space-filling tori and do form torus knots, the simplest of which corresponds to the electromagnetic Hopfion \cite{Arrayas2017, HrideshKedia2017, nastase2019classical}. These solutions have in common the \emph{structure} of their Poynting vector: a Hopf-fibration, moving undistorted with the speed of light \cite{Dalhuisen2014}. Both, the electric and magnetic field lines of the Hopfion also form a Hopf-fibration at one particular moment, before and after which these lines form a tangle of circles, any two of which is linked once. The Poynting vector field of the Hopfion is nowhere vanishing, in contrast to all other solutions corresponding to torus knots. \\
We make use of the spinor formalism \cite{Penrose1984, ODonell2003, TorresdelCastillo2010a} to describe these Maxwell fields and derive a superpotential for them.
We then show that the Sparling-Tod metric is related to the electromagnetic Hopfion. Using the same formalism, we show how the Eguchi-Hanson metric, that, restricted to a particular slice, gives the Eguchi-Hanson instanton \cite{Eguchi1980, Berman2019}, is also related to the Hopfion. In the end, we investigate spin 2 zero rest mass solutions related to linked optical vortices.  
\\
Although some of the said correspondences have appeared in the literature before, these are particular examples of the ones found here.\\ 
\section{Knotted Electromagnetic Fields}\label{section3}
The electromagnetic Hopfion is intimately related to the Robinson congruence in Minkowski space $\mathbb{M}^4$. This congruence is the graphical representation of a non-null twistor \cite{Penrose1967, Penrose1986} and can be obtained by adding a time component to the tangent vector field of the Hopf fibration in such a way that the resulting vector field is null \cite{Dalhuisen2014}. Its spinor form will be denoted by $\Pi^{Rob}$ and is given by\\
\begin{equation}
\Pi^{Rob}\equiv(\Pi^{Rob}_0,\Pi^{Rob}_1)=\tf(x^a)(-t+z-i,x+iy)
\end{equation}
in which $\tf(x^a)$ is an arbitrary scalar function that does not affect the integral curves of the corresponding vector field in $\mathbb{M}^4$. This vector field is obtained with the help of soldering forms\footnote{All implicit conventions used in this work are taken from \cite{Penrose1984}.} (Infeld van der Waerden symbols) and is geodesic and shear free, since $\Pi^{Rob}$ satisfies $\Pi^A\Pi^B\nabla_{BB^\prime}\Pi_A=0$ \cite{Dalhuisen2014}. Any geodesic shear free null congruence can be turned into a null solution of Maxwell's equations such that the anti-self-dual part of the electromagnetic field tensor is given by the Maxwell spinor \cite{Robinson1961}\\
\begin{equation}
\Phi_{AB}=\chi\Pi_A\Pi_B
\end{equation}
\\
A straightforward calculation shows that the corresponding Poynting vector field is \\
\begin{equation}\label{PoyntVect}
\mathbf{S}=2\chi\bar{\chi}k^0\mathbf{k}
\end{equation}
where $(k^0,\mathbf{k})$ corresponds to $\Pi$ and $\bar{\chi}$ denotes the complex conjugate of $\chi$. Therefore, using $\Pi^{Rob}$ leads to a Poynting vector with the structure of a Hopf fibration, irrespective of $\chi$. Although it is possible to find valid $\chi$'s directly, we will use the method of Batemann variables since, by now, literature using these is abundant.\\
\\
Bateman variables, $\alpha, \beta$ are used to construct null electromagnetic fields $\mathbf{F}=\mathbf{\nabla}\alpha\times\mathbf{\nabla}\beta$, with $\mathbf{F}\equiv\mathbf{E}+i\mathbf{B}$ the Riemann-Silberstein vector. Whenever $\alpha, \beta$ satisfy $\mathbf{\nabla}\alpha\times\mathbf{\nabla}\beta=i((\partial_t\alpha)\mathbf{\nabla}\beta-(\partial_t\beta)\mathbf{\nabla}\alpha)$, $\mathbf{F}$ satisfies vacuum Maxwell's equations $\mathbf{\nabla} \centerdot \mathbf{F}=0$ and $\partial_t\mathbf{F}=-i\mathbf{\nabla}\times\mathbf{F}.$ The relation between the Riemann-Silberstein vector and the Maxwell spinor is given by 
\begin{equation}
 \bar{\mathbf{F}}= (\Phi_{00} - \Phi_{11} , i(\Phi_{00} + \Phi_{11} ), -2\Phi_{01})
 \end{equation}

\noindent It is easy to show that we can built new Bateman variables from $\alpha, \beta$ with help of two arbitrary holomorphic functions $f,g: \mathbb{C}^2 \to \mathbb{C}$ if the arguments are the original variables. We then get $\mathbf{F}=\mathbf{\nabla} f(\alpha,\beta) \times \mathbf{\nabla} g(\alpha,\beta)=h(\alpha,\beta)\mathbf{\nabla}\alpha\times\mathbf{\nabla}\beta$, with $h(\alpha,\beta)=\frac{\partial (f,g)} {\partial(\alpha,\beta)}$ the Jacobian of the transformation.
With the choice $f=\alpha^p, g=\beta^q$, where $(p,q)$ are co-prime integers and $\alpha= \frac{r^2-t^2-1+2iz}{r^2-(t-i)^2}, \beta=\frac{2(x+iy)}{r^2-(t-i)^2}$ we do get $\mathbf{F}=pq\alpha^{p-1}\beta^{q-1}\nabla\alpha\times\nabla\beta$, which is the Riemann-Silberstein vector for a null electromagnetic field with field lines in the form of ($p,q$) torus knots \cite{HrideshKedia2017,  Hoyos2015a}. For the Maxwell spinor we then find $\Pi^{Rob}$ with $\tilde{f}\equiv1$ and $\chi=4\frac{pq\alpha^{p-1}\beta^{q-1}}{(r^2-(t+i)^2)^3}$. The Hopfion results for $p=q=1$. In general one can show that for $ \Pi_{A} = \Pi^{Rob} $ with $\tilde{f}\equiv1$
\begin{equation}
\chi = \frac{4\bar{h}}{\bd^3} 
\end{equation}\\
with $\bd = r^2 - (t + i)^2$. It is also possible to construct linked optical vortices, fields that vanish on a closed curve, by a different choice of the holomorphic functions \cite{DeKlerk2017}: 
\begin{equation}\label{OpticalHolomorph}
f(\alpha,\beta)=\int h(\alpha,\beta)d\alpha,\;\; g(\alpha,\beta)=\beta
\end{equation}
with  $\alpha, \beta$ as defined before and $h(\alpha,\beta)$ adjusted to the optical vortex. For example, when ($m,n$) are co-prime we have for a field vanishing on a ($m,n$) torus knot  $h(\alpha,\beta)=\alpha^m+\beta^n$. The method is not limited to torus vortices, cable knot vortices can also be generated. For instance, a $C^{(3,13)}_{(3,2)}$ type cable knot will be found if $h(\alpha,\beta) = \beta^6 - 3\beta^4 \alpha^3 + 3\beta^2 \alpha^6 - 6\beta^2 \alpha^8 - \alpha^9 - 2\alpha^{11}-\alpha^{13}$. Since $\mathbf{F} \propto \mathbf{\nabla}\alpha\times\mathbf{\nabla}\beta$ and we used the Hopfion Bateman variables, we again have the same Poynting vector structure as before.
\section{Superpotential}
In  $\mathbb{M}^4$, we introduce null light-cone coordinates $(u,v, w, \bw)$ and a normalized spinor dyad $((o^A),(\iota^A))$, $\epsilon_{AB}o^A\iota^B=1$, where $\epsilon_{AB}$ is the Levi-Civita symbol, such that the spinor equivalent to the space time point is $X^{AA'}=uo^A\bar{o}^{A^\prime}+v\iota^A\bar{\iota}^{A^\prime} +w o^A\bar{\iota}^{A^\prime}+\bw \iota^A\bar{o}^{A^\prime}$. Let $\delta_{B}\equiv o^{B'}\nabla_{BB^\prime}$.
If $\Phi_{AB}$ is a Maxwell spinor then there exist a superpotential
$\Theta$ such that $\Phi_{AB}=\delta_A\delta_B\Theta$, which can be chosen to satisfy the wave equation as a gauge condition: $\square\Theta=2(\Theta_{,uv}-\Theta_{,w\bw})=0$ \cite{Tod1982}. The Maxwell field is null if and only if  $(\Theta_{,u\bw})^2-\Theta_{,uu}\Theta_{,\bw\bw}=0$, as can be easily verified.\\
\\
The superpotential for the knotted electromagnetic fields discussed in the previous section can be found using the Maxwell spinor form. The result of this excersice is\\
\begin{IEEEeqnarray}{rCl}\label{TorusSuperpot}
\Theta_{(p,q)}  =& &\frac{pq\,(-i)^{q}w^{q-1}}{\sqrt{2}\,(v + \frac{i}{\sqrt{2}})^{q}}\sum_{k = 0}^{\infty} \binom{p-1}{k} \frac{1}{(k+q+1)(k+q)} \frac{1}{\bigg({\frac{i}{\sqrt{2}}\bigg(u+\frac{i}{\sqrt{2}}- \frac{w\bar{w}}{v+\frac{i}{\sqrt{2}}} \bigg)}\bigg)^{k+q}} \nonumber \\
&+& f_1 {u} + f_2 \bw + C  \nonumber \\
\end{IEEEeqnarray}
\\
in which $f_1, f_2$ are functions of $(v,w)$ constrained to $f_{1,v}  = f_{2,w}$ and $C$ is a constant. The null light-cone coordinates $(u,v,w,\bw)$ are related to the Minkowski coordinates $(t,x,y,z)$ by
\begin{equation}\label{nulllightcart}
u  = \frac{1}{\sqrt{2}}(t + z)  ,\; v = \frac{1}{\sqrt{2}}(t - z)  ,\; w = \frac{1}{\sqrt{2}}(x + iy)  ,\; 
\bw = \frac{1}{\sqrt{2}}(x - iy) 
\end{equation}
\\
\section{Anti-Self-Dual Kerr-Schild Spacetime}\label{ASDSec}
In \cite{Tod1982} a construction of half flat spacetimes is presented that is based on the work of Pleba\'{n}ski \cite{Plebanski1975}. The essence of this formalism lies in the fact that null solutions to Maxwell's equations can be associated to solutions of Pleba\'{n}ski's second heavenly equation $\Theta_{,uv} - \Theta_{,w\tw} = (\Theta_{,u\tw})^{2} -\Theta_{,\tw\tw}\Theta_{,uu}$, characterizing a class of asd KS spacetimes. Here the coordinates $(u,v, w, \tw)$ are all complex, in particular $\tw \neq \bw $.
The second heavenly equation defines the following asd solution to Einstein's equation
\begin{IEEEeqnarray}{rCl} \label{ASD}
ds^2&=&2(dudv - dw d\tw) + 2(\Theta_{,uu}dw^{2} + 2\Theta_{,u\tw}dwdv + \Theta_{,\tw\tw}dv^{2})
\end{IEEEeqnarray}
\\
 with the associated Weyl spinor given by $\Psi_{ABCD} = \delta_{A}\delta_{B}\delta_{C}\delta_{D}\Theta = \delta_{A}\delta_{B}\Phi_{CD} $.
It follows that this solution is of KS form if and only if $(\Theta_{,u\tw})^{2} = \Theta_{,\tw\tw}\Theta_{,uu}$, which is indeed the case for the Maxwell fields under consideration.\\
We  have seen that the superpotential for torus knotted electromagnetic fields corresponds to a null Maxwell field and thus, by an appropriate interpretation of the coordinates it is readily seen to satisfy the heavenly equation. On substituting the superpotential of the Hopfion, $\Theta_{(1,1)}$ with $f_{1} = f_{2} = C = 0$, the Sparling-Tod metric is obtained \cite{Jzkkfhlj}, which therefore can be regarded as the simplest one in a class of asd spacetimes: the ($1,1$) member in a family parameterized by ($p,q$). This result can be obtained more directly, without the explicit form of the superpotential.  The metric ($\ref{ASD}$) can be written as \cite{Tod1982}\\
\\
\begin{equation}\label{SpinorASDMetric}
ds^2=(\epsilon_{AB}\epsilon_{A^\prime B^\prime}+\lambda o_{A^\prime}o_{B^\prime}\Phi_{AB})dX^{AA^\prime}dX^{BB^\prime} \;\;\;\;\; 
(\lambda \in \mathbb{C})
\end{equation}\\
where  $\Phi_{AB}$ is the null Maxwell spinor in the Minkowski background. Let us consider the spinor $\Phi_{AB} = \chi \Pi_{A}\Pi_{B} $ where
\begin{equation}\label{torusknot}
(\Pi_A)= \sqrt{2}(-v,w),\;\; \chi = \frac{pq\bigg(1 - \frac{i\sqrt{2}v}{uv - w \tw}\bigg)^{p-1} \bigg(\frac{-\sqrt{2}w}{uv - w\tw}\bigg)^{q-1}}{2(w \tw - uv)^3} 
\end{equation}
 Later we will see that in the Minkowski background this describes the family of torus knotted electromagnetic fields as introduced in section \ref{section3}.
The associated Weyl spinor is of Petrov type $[N]\otimes[-]$:
\begin{IEEEeqnarray}{rCl}\label{Weylpqknot}
		\Psi_{ABCD} = \bigg(\chi \Phi \bigg) \Pi_{A}\Pi_{B}\Pi_{C}\Pi_{D} = \bigg(\frac{\Phi}{\chi}\bigg)  \Phi_{AB} \Phi_{CD}
\end{IEEEeqnarray}
in which $\Phi $ is given by
\begin{IEEEeqnarray}{rCl}
	\Phi = \bigg(M^2 + N\bigg)
\end{IEEEeqnarray}
where M and N are  
\begin{IEEEeqnarray}{rCl}
	M &=& - \Biggr[ \frac{iv(p-1)}{(w\tw - uv)^2 + i\sqrt{2}v(w \tw- uv)}  + \frac{(q-1)}{\sqrt{2}(w\tw - uv)} + \frac{3}{\sqrt{2}(w\tw - uv)}\Biggr] \\
	N &=& \frac{iv(p-1)(iv  + \sqrt{2}(w\tw - uv))}{((w\tw - uv)^2 + i\sqrt{2}v(w \tw- uv))^2} + \frac{(q-1)}{2(w\tw - uv)^2} +\frac{3}{2(w\tw - uv)^2}
\end{IEEEeqnarray}
\noindent Notice that (\ref{Weylpqknot}) is an example of what is known as the `Weyl double copy' \cite{Luna2019} where the key idea is that the Weyl curvature spinor of exact Einstein's equation is related to Maxwell fields in flat spacetime. The theory of the double copy in the classical field theory context consists of a scheme, zeroth copy - single copy - double copy, designating related solutions to respectively a scalar field equation, the abelian Yang-Mills or Maxwell's equations and Einstein's equation \cite{Luna2019, Luna2018}. In the Minkowski background the spinor 
(\ref{torusknot}) corresponds to a Maxwell spinor  with a singularity on the null cone. This can be seen by writing (\ref{torusknot}) in terms of the null light-cone coordinates $(u,v,w,\bw)$. However, since this is a coordinate singularity, it is easily removed via a complex translation $(u,v,w,\bw) \rightarrow (u+ \frac{i}{\sqrt{2}}, v + \frac{i}{\sqrt{2}}, w, \bw) $ \cite{Sparling1981, Tod1981} or in Minkowski coordinates by an imaginary time translation $t \to t + i $.
In doing so, we recover the Robinson congruence. The field so obtained is exactly the family of torus knotted electromagnetic fields described in section \ref{section3} and the associated zero copy is given by the superpotential found in the previous section. 
One could also determine the asd KS spacetime corresponding to a Maxwell field whose zero set forms an algebraic link (section \ref{section3}). The resulting asd KS spacetime would be of type $[N]\otimes[-]$, with a linked optical vortex as single copy and the double copy being it's spin 2 analogue (section \ref{ZeroSetGraviRad}). 

\section{Gravitational Instantons}

Consider the following spinor:

\begin{equation}\label{EguchiMaxwell}
	(\Pi^\prime_A) = \sqrt{2}\big(-\tw,u\big) ,\;\;
	\chi^\prime \,= \frac{pq\bigg({1 - \frac{i\sqrt{2}\tw}{uv- w \tw}}\bigg)^{p-1}{\bigg(\frac{-\sqrt{2}u}{uv - w\tw}\bigg)}^{q-1}}{2(w \tw - u v)^3}  
\end{equation} 

 \noindent It turns out that using this field with $ p = q = 1$  in  (\ref{SpinorASDMetric}) leads to the asd Eguchi-Hanson metric. 
 By taking an Euclidean slice 
 \begin{IEEEeqnarray}{rCl}
 	u = \frac{(t + iz)}{\sqrt{2}},\;\;\;\; v = \bar{u},\;\;\;\; w = \frac{ix - y}{\sqrt{2}},\;\;\;\; \tw = -\bar{w} 
 \end{IEEEeqnarray}
  one obtains the Eguchi-Hanson instanton \cite{Berman2019}. This instanton can therefore be considered the ($1,1$) member in a family parameterized by ($p,q$). 
 The Weyl spinor corresponding to the field in  (\ref{EguchiMaxwell}) is given by\\
\\
\begin{IEEEeqnarray}{rCl}
		\Psi_{ABCD} &=&  \chi^{'}  \bigg[\bigg(L^2 + L_{,\tilde{w}}\bigg) o_{A}o_{B} - 2\bigg(RL + R_{,\tilde{w}}\bigg)\iota_{(A}o_{B)} + \bigg(R^2 + R_{,u}\bigg) \iota_{A}\iota_{B} \bigg]\Pi_{C}^{'}\Pi_{D}^{'} 
\end{IEEEeqnarray}
where  $()$ around indices denotes symmetrization, and 
\begin{IEEEeqnarray}{rCl}
	L &=& - \Biggr[ \frac{i\sqrt{2}uv(p-1)}{(w\tw - uv)^2 + i\sqrt{2}\tw(w \tw- uv)} + \frac{w(q-1)}{w\tw - uv} + \frac{3w}{w\tw - uv}\Biggr]\\
	R &=& \frac{i\sqrt{2}\tw v(p-1)}{(w\tw - uv)^2 + i\sqrt{2}\tw(w \tw- uv)} + \frac{w\tw (q-1)}{u(w\tw - uv)} + \frac{3v}{w \tw- uv}
\end{IEEEeqnarray}\\
 When considered in the Minkowski background, (\ref{EguchiMaxwell}) is a null Maxwell spinor that contains (coordinate) singularities on the light-cone. As before, these can be removed and for $p=q=1$ this leads to a Maxwell field that is, up to sign, the conformal inversion of the Hopfion corresponding to the Sparling-Tod metric. Conformal inversion of an electromagnetic field amounts to $\mathbf{F} \to \bar{\mathbf{F}}(t \to -t)$, and this again is a (different) Hopfion \cite{Arrayas2017, Dalhuisen2014}. Whereas in \cite{Berman2019} a topological trivial single copy of the Eguchi-Hanson instanton was found, we conclude that present field, a Hopfion in Lorentzian spacetime, is a topological non-trivial single copy of the Eguchi-Hanson metric. However, here this metric is a mixed Weyl double copy, the single copy's being a Hopfion and its conformal inverted state. This follows immediately by noting that the Weyl spinor of the Eguchi-Hanson metric is
  $\delta_{A}\delta_{B}(\chi'\Pi_{C}^{'}\Pi_{D}^{'}) = 12(w\tw -uv) \Phi_{AB}^{ST}\Phi_{CD}^{EH}$, a type $[D]\otimes[-]$ with the superscripts denoting the metric.
\section{Gravitational Radiation}\label{ZeroSetGraviRad}
\begin{figure}[h!]
	\centering
	\includegraphics[scale=0.6]{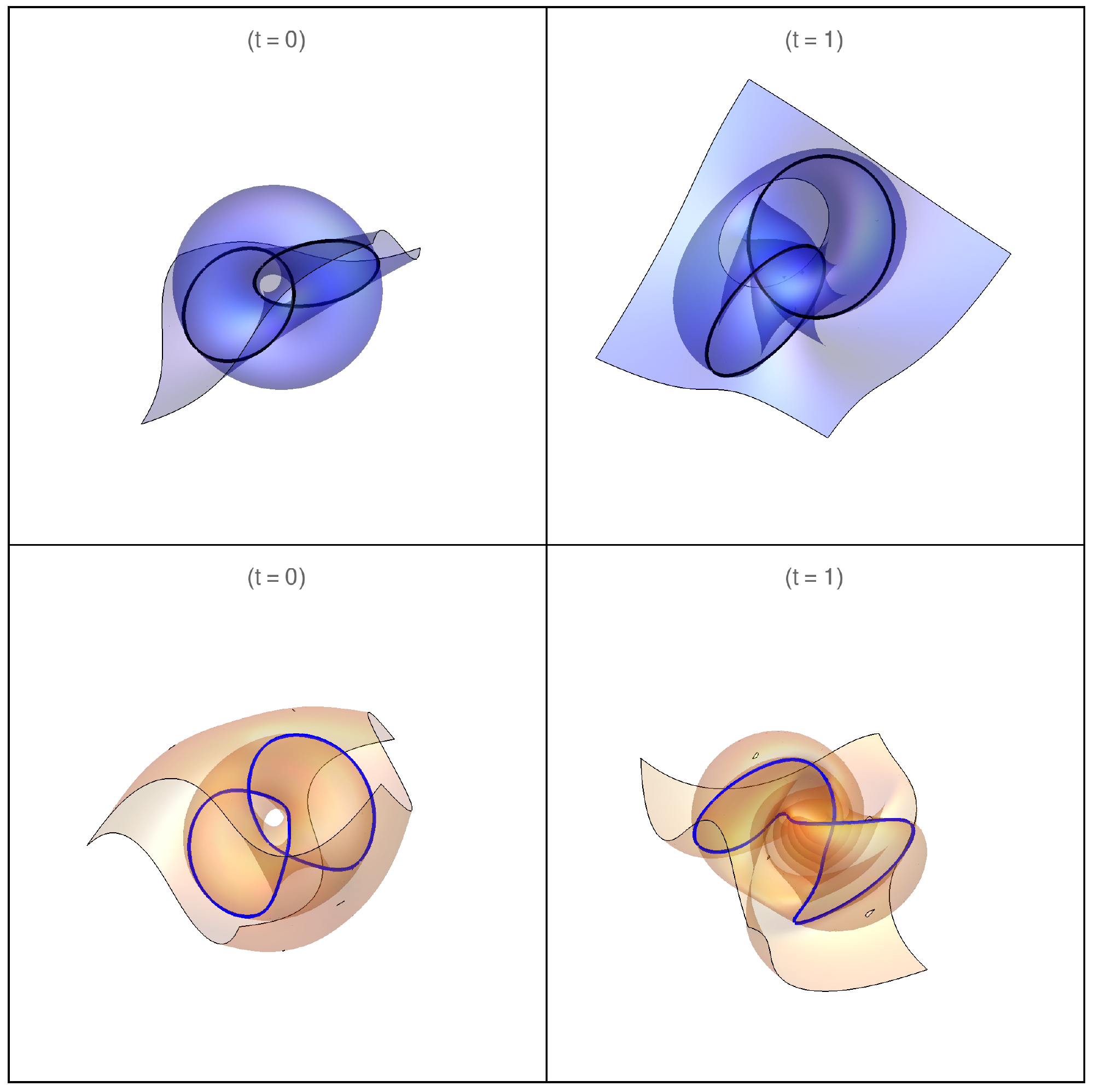}
	\caption{The zeros of the Maxwell (top, in black) and  Weyl (bottom, in blue) spinor obtained by the intersections of the real and imaginary parts of complex $\chi$ (in blue) and  $\mathcal{M}$ (in orange) for $(m = 2,n = 2)$. Plots are shown at $t=0$  and $t=1$ respectively.}
	\label{22VortexZeros}
\end{figure}

In section 4 we used the fact that there must exist a superpotential $\Theta$ for a Maxwell field $\Phi_{AB}=\delta_A\delta_B\Theta$. This is a particular example of a more general statement. The spinor form of Maxwell's equations in vacuum can be written as $\nabla^{AA^\prime}\Phi_{AB}=0$ and it constitutes a spin 1 zero rest mass equation. The spin 2 zero rest mass equation is $\nabla^{AA^\prime}\Phi_{ABCD}=0$ and this represents the linearized Einstein's equation, with $\Phi_{ABCD}$ the Weyl spinor. In general, the solution to the spin n zero rest mass equation $\nabla^{AA^\prime}\Phi_{AB...N}=0$ can be written as $\Phi_{AB...N}=\delta_A\delta_B ...\delta_N\Theta$ in which $\Theta$ is known as the Hertz potential \cite{Penrose1965}. In particular we can write $\Phi_{ABCD}=\delta_A\delta_B\delta_C\delta_D\Theta=\delta_A\delta_B\Phi_{CD}$ and substitute for $\Phi_{CD}$ any of the previously found Maxwell spinors representing torus knotted fields to arrive at a solution of the linearized Einstein's equation. Note that this is reminiscent of the double copy theory in classical field theory context \cite{Luna2018}.
\\
In general, for any electromagnetic field satisfying the two conditions  $\delta_A\Pi_B=0$ and $\delta_A\delta_B\chi=\mathcal{M}\Pi_A\Pi_B$, which is satisfied by all torus knotted fields but also all linked optical vortices considered previously, the corresponding Weyl spinor is given by $\Phi_{ABCD}=\mathcal{M}\Pi_A\Pi_B\Pi_C\Pi_D$, as can be easily checked with help of the spin 1 and spin 2 equations. \\
The Weyl tensor, $C_{abcd}$, associated with this Weyl spinor can be used to define the gravito-electric ($\mathcal{E}_{ij}$) and gravito-magnetic ($\mathcal{B}_{ij}$) fields in the usual way:
\begin{equation}
\mathcal{E}_{ij}=C_{i0j0},\;\;\; \mathcal{B}_{ij}=-*C_{i0j0}
\end{equation}
where $ *$ denotes the Hodge dual operator.
Since $\mathcal{E}_{ij}\xi^j$ measures the relative (tidal) acceleration between two points separated by a small vector $\bm{\xi}$, we can interpret the electric part of the Weyl tensor as a tidal field. The matrix $(\mathcal{E}_{ij})$ is traceless and symmetric, and can therefore be characterized by its eigenvalues and eigenvectors, the integral curves of which are called tendex lines \cite{Nichols2011} and can be considered the gravitational analogue of electric field lines. An extended object placed in the field $(\mathcal{E}_{ij})$ will be stretched in the direction of the lines corresponding to positive eigenvalues and compressed along the curves corresponding to negative eigenvalues. The strength of this effect is related to the eigenvalue. The interpretation of the magnetic part of the Weyl tensor is as a frame-drag field. A gyroscope at the tip of the vector $\bm{\xi}$ will precess with angular velocity $\mathcal{B}_{ij}\xi^j$ relative to inertial frames at the tail. The matrix $(\mathcal{B}_{ij})$ is also traceless and symmetric and the integral curves of the eigenvectors are now called vortex lines \cite{Nichols2011, Maartens1998}.\\
\\
 For the fields encapsulated by the two conditions, we find that the eigenvalues of both the gravito-electric and the gravito-magnetic fields are: 
 \begin{equation}\label{Eigenval}
 -\frac{1}{2}(\mathcal{M}\bar{\mathcal{M}})^{\frac{1}{2}}(\Pi_{0}\Pi^{\dagger}_{0} + \Pi_{1}\Pi^{\dagger}_{1})^2,\;\; 0 \;,\;\;
  \frac{1}{2}(\mathcal{M}\bar{\mathcal{M}})^{\frac{1}{2}}(\Pi_{0}\Pi^{\dagger}_{0} + \Pi_{1}\Pi^{\dagger}_{1})^2 
 \end{equation}
The associated gravito-electric eigenvectors are $\bm{\mathcal{EG}_{-}}, \bm{\mathcal{EG}_{0}}$ and  $\bm{\mathcal{EG}_{+}}$, the gravito-magnetic ones are $\bm{\mathcal{BG}_{-}}, \bm{\mathcal{BG}_{0}}$ and  $\bm{\mathcal{BG}_{+}}$ with\\
\\
\begin{IEEEeqnarray}{rCl}\label{ZeroEigen}
\bm{\mathcal{EG}_{0}}=\bm{\mathcal{BG}_{0}}  = \begin{pmatrix}
&-\bigg(\Pi_{1}\Pi^{\dagger}_{0} + \Pi_{0}\Pi^{\dagger}_{1}\bigg)&\\
&i \bigg(\Pi_{1}\Pi^{\dagger}_{0} - \Pi_{0}\Pi^{\dagger}_{1}\bigg)&\\
&\bigg(-\Pi_{0}\Pi^{\dagger}_{0} + \Pi_{1}\Pi^{\dagger}_{1}\bigg)&
\end{pmatrix}\propto \mathbf{S}
\end{IEEEeqnarray}
where $\mathbf{S}$ is the Poynting vector field (\ref{PoyntVect}).\\
\begin{IEEEeqnarray}{rCl}
\bm{\mathcal{EG}_{+}}=\begin{pmatrix}
&\bigg((\mathcal{M})^{\frac{1}{2}}(-\Pi_{0}^2 + \Pi_{1}^{2}) + (\bar{\mathcal{M}})^{\frac{1}{2}} (-\Pi^{\dagger 2}_{0} + \Pi_{1}^{\dagger 2})\bigg)&\\
&-i \bigg((\mathcal{M})^{\frac{1}{2}}(\Pi_{0}^2 + \Pi_{1}^{2}) - (\bar{\mathcal{M}})^{\frac{1}{2}} (\Pi^{\dagger 2}_{0} + \Pi_{1}^{\dagger 2})\bigg)&\\
&2 \bigg((\mathcal{M})^{\frac{1}{2}}\Pi_{0}\Pi_{1} + (\bar{\mathcal{M}})^{\frac{1}{2}} \Pi^{\dagger}_{0}\Pi^{\dagger}_{1}\bigg)&
\end{pmatrix}
\end{IEEEeqnarray}\\
\begin{IEEEeqnarray}{rCl}
\bm{\mathcal{EG}_{-}}=\begin{pmatrix}
&-i\bigg((\mathcal{M})^{\frac{1}{2}}(-\Pi_{0}^2 + \Pi_{1}^{2}) + (\bar{\mathcal{M}})^{\frac{1}{2}} (\Pi^{\dagger 2}_{0} - \Pi_{1}^{\dagger 2})\bigg)&\\
&-\bigg((\mathcal{M})^{\frac{1}{2}}(\Pi_{0}^2 + \Pi_{1}^{2}) + (\bar{\mathcal{M}})^{\frac{1}{2}} (\Pi^{\dagger 2}_{0} + \Pi_{1}^{\dagger 2})\bigg)&\\
&-2i\bigg((\mathcal{M})^{\frac{1}{2}}\Pi_{0}\Pi_{1} - (\bar{\mathcal{M}})^{\frac{1}{2}} \Pi^{\dagger}_{0}\Pi^{\dagger}_{1}\bigg)&
\end{pmatrix}
\end{IEEEeqnarray}\\
\begin{IEEEeqnarray}{rCl}
\bm{\mathcal{BG}_{-}}=(1-i)\begin{pmatrix}
&\bigg((\mathcal{M})^{\frac{1}{2}}(-\Pi_{0}^2 + \Pi_{1}^{2}) - i (\bar{\mathcal{M}})^{\frac{1}{2}} (\Pi^{\dagger 2}_{0} - \Pi_{1}^{\dagger 2})\bigg)&\\
&- \bigg(i (\mathcal{M})^{\frac{1}{2}}(\Pi_{0}^2 + \Pi_{1}^{2}) + (\bar{\mathcal{M}})^{\frac{1}{2}} (\Pi^{\dagger 2}_{0} + \Pi_{1}^{\dagger 2})\bigg)&\\
&2\bigg((\mathcal{M})^{\frac{1}{2}}\Pi_{0}\Pi_{1} + i (\bar{\mathcal{M}})^{\frac{1}{2}} \Pi^{\dagger}_{0}\Pi^{\dagger}_{1}\bigg)&
\end{pmatrix}
\end{IEEEeqnarray}\\
\begin{IEEEeqnarray}{rCl}\label{BGminEigenvec}
\bm{\mathcal{BG}_{+}}=(1+i)\begin{pmatrix}
&\bigg((\mathcal{M})^{\frac{1}{2}}(-\Pi_{0}^2 + \Pi_{1}^{2}) + i (\bar{\mathcal{M}})^{\frac{1}{2}} (\Pi^{\dagger 2}_{0} - \Pi_{1}^{\dagger 2})\bigg)&\\
& \bigg(-i (\mathcal{M})^{\frac{1}{2}}(\Pi_{0}^2 + \Pi_{1}^{2}) + (\bar{\mathcal{M}})^{\frac{1}{2}} (\Pi^{\dagger 2}_{0} + \Pi_{1}^{\dagger 2})\bigg)&\\
&2\bigg((\mathcal{M})^{\frac{1}{2}}\Pi_{0}\Pi_{1} - i (\bar{\mathcal{M}})^{\frac{1}{2}} \Pi^{\dagger}_{0}\Pi^{\dagger}_{1}\bigg)&
\end{pmatrix}
\end{IEEEeqnarray}\\~\\
  Since all type N solutions of linearized Einstein's equation can be written in the form $\Phi_{ABCD}=\mathcal{M}\Pi_A\Pi_B\Pi_C\Pi_D$, the expressions above for the eigenvalues and eigenvectors have  applicability beyond what is explored here.
 By defining $\bm{\mathcal{F_{E}}} =\bm{\mathcal{EG}_{-}} + i \bm{\mathcal{EG}_{+}}$ and $\bm{\mathcal{F_{B}}} = \bm{\mathcal{BG}_{-}} + i \bm{\mathcal{BG}_{+}}$, we find 
$\bm{\mathcal{F_{E}}} = \frac{e^{\frac{i\pi}{4}}}{\sqrt{2}} \bm{\mathcal{F_{B}}} = \big(\frac{-2 i (\bar{\mathcal{M}})^{\frac{1}{2}}}{\bar{\chi}}\big)\mathbf{F}$, where  $\mathbf{F}$ is the Riemann-Silberstein vector of the corresponding electromagnetic field. These results can be seen as a generalization of the ones stated in \cite{Dalhuisen2014, Thompson2014a, Thompson2015}.
\\
The equations above make it clear that the gravitational radiation that corresponds to any of the electromagnetic fields defined by the two conditions above, which contain ($p,q$) torus knotted fields and linked optical vortices, posses a Hopf structure as the curves along which no stretching, compression or precession will occur. \\~\\
Let us consider the case of linked optical vortices, where the principal spinor is $ \Pi_{A} = \Pi^{Rob}$ and $ \tilde{f} \equiv 1$.
For a ($m,n$) optical vortex, one finds the following complex $\mathcal{M}$ function of the Weyl spinor:
\begin{IEEEeqnarray}{rCl}
	\mathcal{M}
 = \frac{16}{\bd^8}\bigg[\ba^m\bd^3(6 - i \sqrt{2} \ba^{-2}\vp m(m-1))+ i \ba^{m-1}m\vp(5\sqrt{2}\bd^2 - 12w) + \frac{\bb^{n}\bd^{3}}{2} (12 + 7n + n^2) \bigg]\nonumber \\
\end{IEEEeqnarray}
with $\ba$, $ \bb$, $\bd$, $v$, $w$ as defined in section 3 and 4, $ \vp = v + \frac{i}{\sqrt{2}}$ and ($m,n$) integers.\\
As an example we take $m=2, n=2$ and consider the points where the Weyl tensor vanishes, $\mathcal{M} \equiv 0$. The ($2,2$) optical vortex contains a Hopf link where the field vanishes. The same is true for the Weyl tensor. However, whereas in the electromagnetic case the topology of this link is conserved, in linearized gravity this link deforms into an unknot (fig. \ref{22VortexZeros}).

\begin{figure}[h!]
	\centering
	\includegraphics[scale=0.6]{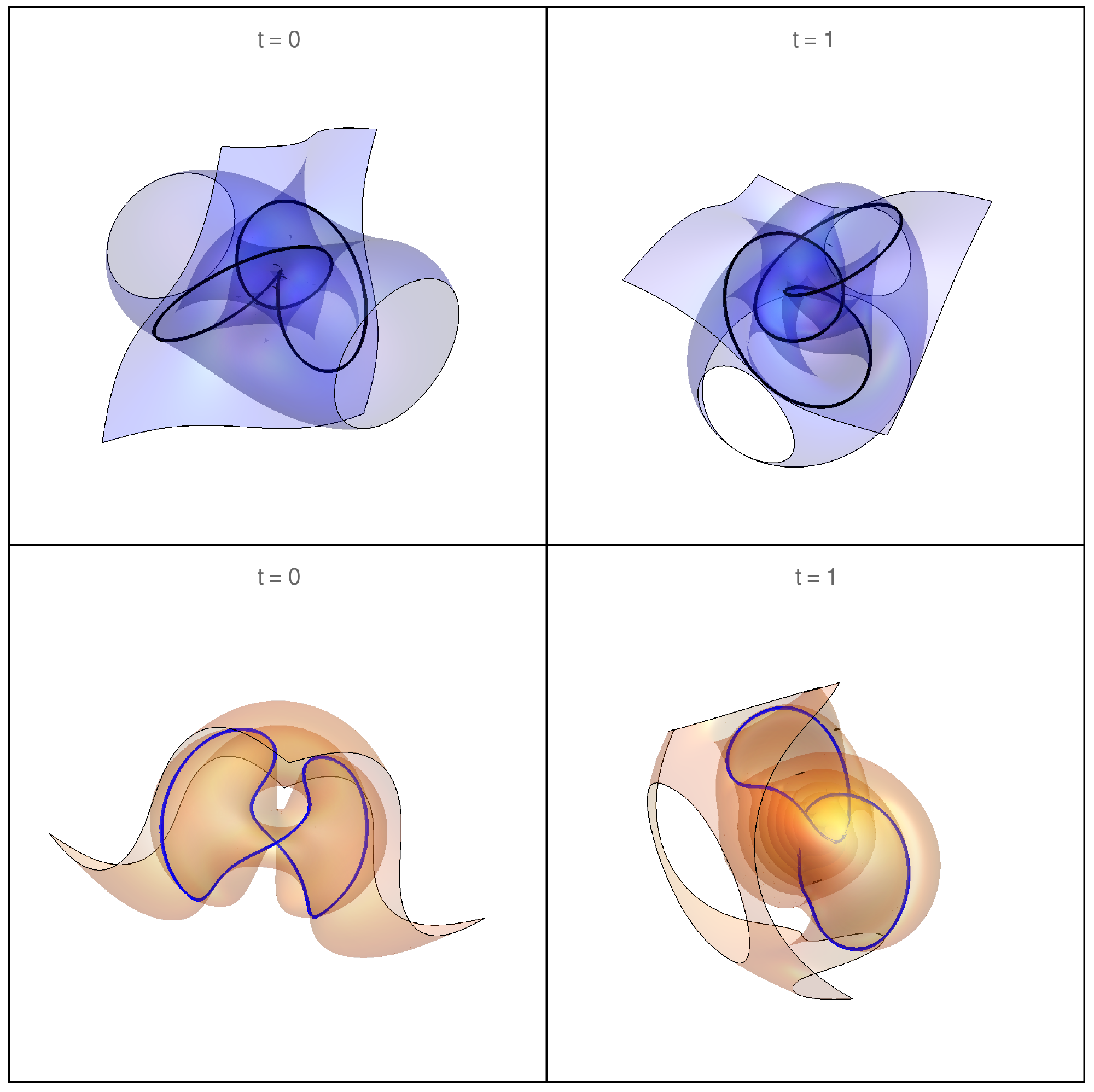}
	\caption{The zeros of the Maxwell (top, in black) and  Weyl (bottom, in blue) spinor obtained by the intersections of the real and imaginary parts of the complex $\chi$ (in blue) and  $\mathcal{M}$ (in orange) for $(m = 3,n = 2)$. Plots are shown at $t=0$  and $t=1$ respectively.}
	\label{32VortexZeros}
\end{figure}

\noindent As a second example we take $m=3$ and $n=2$. In general, for ($m,n$) co-prime and $m,n > 1$, the optical vortex vanishes on a ($m,n$) torus knot and this does not change in time. Again, in linearized gravity the zero set does change: an unknot is deformed into a Hopf link (fig. \ref{32VortexZeros}). 

\begin{figure}[h!]
	\centering
	\includegraphics[scale=0.5]{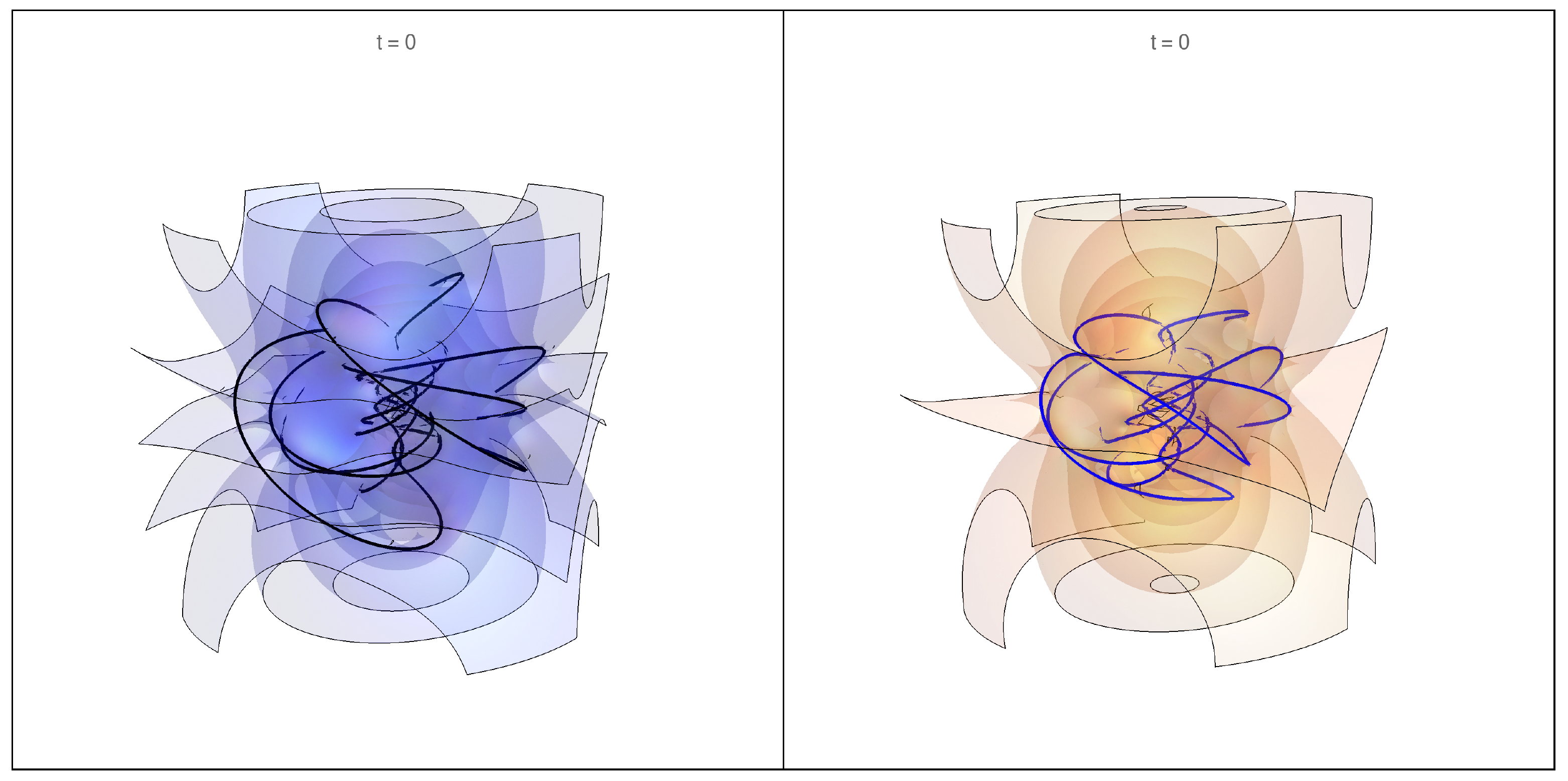}
	\caption{The zeros of the Maxwell (left, in black) and Weyl (right, in blue) spinor obtained by the intersections of the real and imaginary parts of the complex $\chi$ (in blue) and  $\mathcal{M}$ (in orange) for a $C^{(3,13)}_{(3,2)}$ cable knot. Plots are shown at $t=0$.}
	\label{CableVortexZeros}
\end{figure}

\noindent Finally, for a $C^{(3,13)}_{(3,2)}$ type cable knot vortex, the complex $ \mathcal{M}$ function is given by
\begin{IEEEeqnarray}{rCl}
\mathcal{M} &=& \frac{-16}{\bd^8} \Bigg\{-12 i \ba^2 \vp w \bigg[(\ba^2+1)(13 \ba^2+9) \ba^6+6 (8
	\ba^2-3) \ba^3 \bb^2+9 \bb^4\bigg] +\sqrt{2} \bd^2\bigg[-48 \bb w \nonumber \\
	&(&-2 \ba^8+\ba^6-2 \ba^3 \bb^2+\bb^4)+ 5 i \ba^2 \vp
	((\ba^2+1) (13 \ba^2+9) \ba^6+6 (8 \ba^2-3) \ba^3
	\bb^2+9 \bb^4)\bigg] \nonumber \\
	&+&2 \bd^3 \bigg[9 \ba \bb^4(\ba^2-i \sqrt{2}
	\vp)+3 \ba^4 \bb^2 (6 \ba^4+\ba^2 (-3-56 i \sqrt{2} \vp)+15 i \sqrt{2}
	\vp) + \nonumber \\
	\;\;\;&&\ba^7 (3 (\ba^3+\ba)^2-2 i \sqrt{2} (39 \ba^4+55
	\ba^2+18) \vp)-3 \bb^6\bigg] +24 \bd \bigg[w^2 (2 \ba^8-\ba^6+6
	\ba^3 \bb^2-5 \bb^4) \nonumber \\
	&-& 2 \ba^2 \bb \vp^2 (8 \ba^5-3 \ba^3+3
	\bb^2)+2 i \ba^2 \bb \vp w (8 \ba^5-3 \ba^3+3
	\bb^2)\bigg]\Bigg\}\nonumber \\
\end{IEEEeqnarray}
\noindent In fig. (\ref{CableVortexZeros}), we observe that the zeros of the Weyl tensor do correspond to a $C^{(3,13)}_{(3,2)}$ type cable knot just as its electromagnetic counterpart. 
\\
\noindent Various other values for ($m, n$) have been investigated numerically. In the electromagnetic case these correspond to vortices in the following way.
The topology of the vortex for
\begin{enumerate}
	\item $m = 1$ or $n = 1$: unknot.
	\item $m>1$, $n>1$: \begin{itemize}
								\item and co-prime: ($m, n$) torus knot.
								\item not co-prime: \begin{itemize}
														\item $ m=n$: set of $n$ linked circles.
														\item $m\neq n$: $gcd(m,n)$ linked copies of a ($\tilde{m},\tilde{n} $) torus knot with $\tilde{m} = \frac{m}{gcd(m,n)} $, $\tilde{n} = \frac{n}{gcd(m,n)} $ and $gcd=$ greatest common divisor.
								\end{itemize}
	  	\end{itemize}
\end{enumerate}
\noindent For the zero set of the corresponding Weyl tensor we found the following.
\begin{enumerate}
	\item $m=2, n$ arbitrary: for $t=0$ the topology matches the Maxwell field vortex, but it eventually becomes an unknot.
	\item $m=3, n$ arbitrary: the topology deforms from an unknot to a Maxwell ($2, n$) vortex and finally back to an unknot. 
	\item $m > 3, n$ arbitrary: the topology seems to be conserved in time and is the same as the Maxwell ($m-2,n$) vortex.
\end{enumerate}
\noindent Since the Weyl spinors obtained in the previous sections can be considered solutions of complex linearized Einstein's equation, one can easily obtain real solutions of linearized Einstein's equation, by taking a Lorentzian slice.
 \begin{equation}
 u = \frac{(t + z)}{\sqrt{2}},\;\;\;\; v = \frac{(t-z)}{\sqrt{2}}, \;\;\;\; w = \frac{x + i y}{\sqrt{2}}, \;\;\;\; \tw = \bw
 \end{equation}

\noindent The singularity is removed by the complex translation of the time coordinate $t \rightarrow t + i $, exactly as in the case of null Maxwell fields. For the Weyl spinor in  (\ref{Weylpqknot}), we would obtain a type N knotted solution of linearized Einstein's equation in vacuum. This solution has been investigated in \cite{Thompson2014a} by making use of the Penrose transform. Likewise one could reformulate the solutions of spin 1 and spin 2 zero mass equations to construct and characterize an asd spacetime. 
For this, we have to allow the null light-cone coordinates, in which the  spin 1 and spin 2 fields can be described, to become complex ($u,v,w,\tw$) with $w$ and $\tw$ unrelated. The resulting complex Maxwell field can be plugged into (\ref{SpinorASDMetric}) to determine the metric and the spin 2 solution following this reinterpretation can be used to classify the Petrov type of this metric.

\section{Summary}
We found superpotentials for electromagnetic knots that satisfy Pleba\'{n}ski's second heavenly equation and defines anti-self-dual solutions to Einstein's equation in Kerr-Schild form. The Sparling-Tod metric is among the so defined solutions and relates to the Hopfion, the ($1,1$) member of a family of ($p,q$) torus knotted fields. Therefore, the Sparling-Tod metric is the ($1,1$) member of a family of spacetimes parameterized by ($p,q$). This result is also obtained more directly from the spinor form of the electromagnetic field. The last method is used in relation to the Eguchi-Hanson metric, which is also found to be the ($1,1$) member of a family parameterized by ($p,q$). By taking Euclidean slices, this parameterization carries over to a family of instantons, of which the Eguchi-Hanson instanton is the ($1,1$) member. These results are discussed in light of the theory of the Weyl double copy. It is found that the Sparling-Tod metric is the pure double copy of the Hopfion and the Eguchi-Hanson metric the mixed double copy of the same Hopfion and its conformally inverted field. We considered the zero rest mass equation for spin 1 (electromagnetism) and spin 2 (linearized gravity) and derived general expressions for eigenvalues and eigenvectors of the gravito-electric and gravito-magnetic parts of the Weyl tensor corresponding to  electromagnetic fields that satisfy two conditions. All torus knotted fields and optical vortices satisfy these conditions. It is found that all solutions to linearized gravity do possess a Hopf structure as the curves along which no compression, stretching or precession will occur. We focussed attention on three examples, two torus vortices, and a cable knot vortex, and plotted the zeros of the Weyl tensor for these. We did investigate many other members of the ($m,n$) linked optical vortices and their corresponding linearized gravity solutions numerically. It was found that for small $m$ the
zero set of the Weyl tensor changes topology, whereas for $m>3$ the topology seems to be conserved. Besides, in the latter case, there is a direct correspondence with the ($m-2,n$) electromagnetic vortex. Although gravitational fields that correspond to linked optical vortices have here only been considered in the linearized regime, the Maxwell spinor for vortices can also be used as input in the formalism described in section 5 to derive the associated anti-self-dual Kerr-Schild spacetimes.\\

\newpage
\providecommand{\href}[2]{#2}\begingroup\raggedright\endgroup
\end{document}